\documentclass[onecolumn,preprintnumbers,amsmath,amssymb,notitlepage]{revtex4}
\usepackage{graphicx}
\usepackage{epsfig}
\usepackage{amsmath}
\usepackage{amssymb}
\usepackage{dcolumn}

\begin{document}

\title{Generalized Quantization Principle in Canonical Quantum Gravity and Application to Quantum Cosmology}

\author{Martin Kober}
\email{kober@fias.uni-frankfurt.de}
 
\affiliation{Frankfurt Institute for Advanced Studies (FIAS),
Johann Wolfgang Goethe-Universit\"at,
Ruth-Moufang-Strasse 1, 60438 Frankfurt am Main, Germany}

\date{\today}

\begin{abstract}
In this paper, a generalized quantization principle for the gravitational field in canonical
quantum gravity, especially with respect to quantum geometrodynamics is considered. This
assumption can be interpreted as a transfer from the generalized uncertainty principle in
quantum mechanics, which is postulated as generalization of the Heisenberg algebra to
introduce a minimal length, to a corresponding quantization principle concerning the
quantities of quantum gravity. According to this presupposition there have to be given
generalized representations of the operators referring to the observables in the canonical
approach of a quantum description of general relativity. This also leads to generalized
constraints for the states and thus to a generalized Wheeler–DeWitt equation determining
a new dynamical behaviour. As a special manifestation of this modified canonical theory
of quantum gravity, quantum cosmology is explored. The generalized cosmological Wheeler–DeWitt
equation corresponding to the application of the generalized quantization principle to the
cosmological degree of freedom is solved by using Sommerfelds polynomial method.
\end{abstract}

\maketitle

\section{Introduction}

The search for a quantum theory of gravity is one of the most interesting if not the most interesting
research topic in contemporary fundamental physics. A very important approach to a quantum description
of general relativity is canonical quantum gravity, where are postulated canonical commutation relations
for the gravitational field and its canonical conjugated quantity. Especially in its manifestation
as loop quantum gravity \cite{Rovelli:1989za},\cite{Rovelli:1994ge} based on the new formulation of
canonical general relativity given in \cite{Ashtekar:1986yd},\cite{Ashtekar:1987gu} it is considered
as a very promising candidate for a quantum theory of gravity.
Another important concept in the context of quantum gravity is the so called generalized uncertainty principle,
which refers to quantum mechanics and consists in the assumption of a generalization of the usual Heisenbergian
commutator between position and momentum. The generalized uncertainty principle, which is closely related to the
concept of noncommutative geometry, is a well-known concept and introduces the existence of a fundamental minimal
length to quantum mechanics and quantum field theory. This kind of a generalized uncertainty relation and the
corresponding introduction of a minimal length has been explored in \cite{Maggiore:1993kv},\cite{Maggiore:1993rv},\cite{Maggiore:1993zu},\cite{Kempf:1994su},\cite{Hinrichsen:1995mf},\cite{Kempf:1996ss},\cite{Kempf:1996nk} for the first time. 
Usually the generalized uncertainty principle is interpreted as an effective description of consequences
of a fundamental quantum theory of gravity, since the existence of a minimal length in the Planck regime
is assumed to be an implication of a quantum description of the gravitational field. However, there exists
also the possibility to consider the generalized uncertainty principle as a fundamental description of
nature by itself. If this would be the case, then one should assume that not only the variables of quantum
mechanics, the position and momentum of a particle, obey such a generalized uncertainty principle, but also
other variables have to be quantized by using a generalized quantization rule.
In the common description of quantum field theory and canonical quantum gravity the usual Heisenbergian
commutation relation between the position and momentum operator is transferred to the field and its canonical
conjugated quantity. If there is postulated a generalized uncertainty principle in quantum mechanics as fundamental
description of nature, then the transfer to quantum field theory and quantum gravity should be performed as well.
In \cite{Matsuo:2005fb} there has been treated the quantization of fields based on the generalized uncertainty principle.
Although the relation of the generalized uncertainty principle and the corresponding minimal length to gravity
has been studied in many papers
\cite{Scardigli:1999jh},\cite{Capozziello:1999wx},\cite{Crowell:1999up},\cite{Chang:2001bm},\cite{Kim:2006rx},\cite{Park:2007az},\cite{Scardigli:2007bw},\cite{Kim:2007hf},\cite{Bina:2007wj},\cite{Battisti:2007zg},\cite{Zhu:2008cg},\cite{Battisti:2008rv},\cite{Vakili:2008zg},\cite{Myung:2009gv},\cite{Myung:2009ur},\cite{Ali:2009zq},\cite{Farmany:2009zz},\cite{Li:2009zz},\cite{Bina:2010ir},\cite{Kim:2010wc},\cite{Ali:2011ap},\cite{Das:2008kaa},\cite{Ashoorioon:2004vm},\cite{Ashoorioon:2004rs},\cite{Ashoorioon:2004wd},\cite{Ashoorioon:2005ep},\cite{Vakili:2008tt},\cite{Kim:2007bx},
\cite{Garattini:2010dn},
it seems that a modification of the quantization principle of the gravitational field itself has been omitted so far.
In \cite{Majumder:2011ad} and \cite{Majumder:2011bv} has already been studied the generalized uncertainty principle
with respect to the gravitational degrees of freedom of special scenarios of a black hole and cosmology.
But in these papers this transfer of the generalized uncertainty principle to gravitational degrees of freedom
seems not to be considered as a special manifestation of a fundamental general description of canonical quantum
gravity. At least there is given no general formulation.
In the present paper is hold the assumption that a generalized uncertainty principle with respect to the
gravitational field could represent a fundamental description of nature and it is considered within
the description of quantum geometrodynamics. But of course it can analogously be transferred to the
quantization procedure within loop quantum gravity. The extended quantum description of the gravitational
field is applied to quantum cosmology as special manifestation of the general theory.  
In \cite{Kober:2010sj} has been considered the gauge description of general relativity under incorporation
of a generalized uncertainty principle for quantum mechanics. In this description the generalized uncertainty
principle already influences the fundamental description of gravity decisively. The theory has been considered
as a kind of semiclassical description. In the theory of this paper the generalized uncertainty principle now
refers to the fundamental quantum description of general relativity itself. A generalized quantization principle
for the gravitational field can under certain conditions imply nontrivial commutation relations for the components
of the gravitational field in the same way as a generalized uncertainty principle in quantum mechanics can imply
noncommutative geometry. Such nontrivial commutation relations for the components of the gravitational field
have been explored already in \cite{Kober:2011am}.
   
The paper is structured as follows: First there is given a short repetition of the generalized uncertainty principle
in usual quantum mechanics. Then the generalized uncertainty principle is transferred to the gravitational
field and accordingly it is introduced a generalized quantization principle for the variables of canonical quantum gravity.
There are considered two cases: In the first case there is assumed that the commutator between the variables of canonical
quantum gravity depends on the canonical conjugated quantity and in the second case there is assumed that the commutator
depends on the gravitational field itself. There are given the corresponding representations of the operators and the
corresponding generalized uncertainty relations. After this there are derived the generalized constraints on the quantum
state of the gravitational field and thus there are obtained the generalized Wheeler-DeWitt equations. For the first
case of the generalized quantization principle, which commutator depends on the canonical conjugated quantity to the
gravitational field, is considered the application to the generalized dynamics of quantum cosmology. The corresponding
Wheeler-DeWitt equation is obtained by applying the generalized quantization principle for the gravitational field
to the cosmological degree of freedom, which is contained in the scale factor under presupposition of a Friedmann
Lemaitre universe. It is found a solution by using Sommerfelds polynomial approach, which yields a recursion condition
for the coefficients of the polynomial presupposed as solution ansatz for the quantum state depending on the
cosmological scale factor by inserting a general polynomial to the generalized cosmological Wheeler-DeWitt
equation. The obtained recursion condition maintains that the corresponding state converges, which means
that it is normalizable and can thus be considered as physical solution.

\section{Generalized Quantization Principle for the Gravitational Field}

\subsection{Generalized Uncertainty Principle in Quantum Mechanics}

Usually a generalized uncertainty principle is assumed with respect to the variables of quantum mechanics.
The Heisenbergian commutation relation, $[\hat x^\mu,\hat p_\nu]=i\delta^{\mu}_\nu$, is extended in such a way that
the commutator depends on the position or on the momentum,

\begin{equation}
[\hat x^\mu,\hat p_\nu]=i\left\{\delta^{\mu}_\nu\left[1+\lambda a(\hat x,\hat p)\right]
+\kappa b^{\mu}_{\nu}(\hat x,\hat p)\right\},
\label{generalized_algebra_quantum_mechanics}
\end{equation}
where $a(\hat x,\hat p)$ and $b^\mu_\nu(\hat x,\hat p)$ describe two arbitrary functions depending on the
position and the momentum and $\lambda$ and $\kappa$ are constants. For the special case of

\begin{equation}
[\hat x^\mu,\hat p_\nu]=i\left\{\delta^{\mu}_\nu \left[1+\lambda \hat p^\rho \hat p_\rho\right]
+2\kappa \hat p^\mu \hat p_\nu\right\},\quad \kappa=\lambda,
\label{special_generalied_algebra_quantum_mechanics}
\end{equation}
this leads to the following generalized uncertainty relation:

\begin{eqnarray}
&&\Delta x^\mu \Delta p_\mu \geq \langle \psi|[\hat x^\mu,\hat p_\mu]|\psi\rangle=
\frac{1}{2}\left[1+3\lambda \Delta p^\mu \Delta p_\mu
+3\lambda \langle \hat p^\mu \rangle \langle \hat p_\mu \rangle\right]\quad {\rm if}\quad \mu=\nu,\nonumber\\
&&\Delta x^\mu \Delta p_\nu \geq \langle \psi|[\hat x^\mu,\hat p_\nu]|\psi\rangle=
\lambda \langle \hat p^\mu \rangle \langle \hat p_\nu \rangle\quad {\rm if}\quad \mu \neq \nu. 
\label{generalized_uncertainty_quantum_mechanics}
\end{eqnarray}
The special case ($\ref{special_generalied_algebra_quantum_mechanics}$) for the generalized uncertainty principle
leads to a quite simple representation of the position and momentum operator in position space, where has to be
performed a series expansion and which is given by

\begin{eqnarray}
\hat x^\mu |\psi(x)\rangle=x^\mu |\psi(x)\rangle \quad,\quad \hat p_\mu |\psi(x)\rangle
=-i\left(1-\lambda \partial^\rho \partial_\rho\right)\partial_\mu |\psi(x)\rangle
+\mathcal{O}\left(\lambda^2\right),
\label{generalized_operators_quantum_mechanics_position}
\end{eqnarray}
whereas its corresponding representation in momentum space is given by

\begin{eqnarray}
\hat x^\mu |\psi(p)\rangle=\left[i\left(1+\lambda p^\rho p_\rho\right)\partial_p^\mu
+2i\lambda p^\mu p_\rho \partial_p^\rho\right]|\psi(p)\rangle
\quad,\quad \hat p_\mu |\psi(p)\rangle=p_\mu |\psi(p)\rangle.
\label{generalized_operators_quantum_mechanics_momentum}
\end{eqnarray}
In this paper the generalized uncertainty principle ($\ref{generalized_algebra_quantum_mechanics}$) is considered
as a fundamental description of nature and accordingly it should be transferred to the variables of canonical
quantum gravity, what is done with respect to the formulation of quantum geometrodynamics in this paper.

\subsection{Transfer to a Generalized Quantization Principle for the Gravitational Field}

If a generalized uncertainty principle with respect to quantum mechanics ($\ref{generalized_algebra_quantum_mechanics}$)
is not interpreted as an indirect consequence of a fundamental quantum theory of gravity and thus as an effective description
but as a fundamental property of quantum theory, then one should expect that it has to be transferred to field quantization.
Usually the Heisenbergian commutation relation of usual quantum mechanics, $[\hat x_\mu,\hat p^\nu]=i\delta_\mu^\nu$, is
considered as fundamental postulate to obtain operators and quantum states from the quantities of classical mechanics.
This quantization principle is transferred to the quantum description of fields leading to the commutation relation:
$[\hat \phi(x),\hat \pi_\phi(y)]=i\delta(x-y)$, if $\hat \phi(x)$ denotes a scalar field and $\hat \pi_\phi(x)$ denotes its
canonical conjugated quantity. If ($\ref{generalized_algebra_quantum_mechanics}$) is now postulated as fundamental
commutation relation of quantum mechanics and it is transferred to field quantization, this leads to: $[\hat \phi(x),\hat \pi_\phi(y)]=i\delta(x-y)\left[1+\lambda A(\hat \phi,\hat \pi_\phi)\right]$. But if the generalized quantization principle
is transferred to fields then one should expect that it also holds for the special case of the gravitational field.
To obtain the analogue quantization rule for the gravitational field one has of course to refer to the canonical
quantum description of general relativity. In the canonical formulation of general relativity space-time is foliated
into a spacelike hypersurface and a time coordinate leading to the following representation of the metric tensor:

\begin{equation}
g_{\mu\nu}=\left(\begin{matrix}N_a N^a-N^2 & N_b\\ N_c & h_{ab}\end{matrix}\right),
\end{equation}
where $h_{ab}$ denotes the three metric on the spacelike hypersurface, $N$ denotes the lapse function
and $N^a$ is the shift vector. If there is used this representation of the metric tensor, then the
Einstein Hilbert action reads as follows \cite{Kiefer:2004}:

\begin{eqnarray}
S_{EH}&=&\int_{\mathcal{M}}dt\ d^3 x\ \mathcal{L}_g=\frac{1}{16 \pi G}\int_{\mathcal{M}}dt\ d^3 x\ N
\left(G^{abcd}K_{ab}K_{cd}+\sqrt{h}\left[R_h-2\Lambda\right]\right)\nonumber\\
&=&\frac{1}{16 \pi G}\int_{\mathcal{M}}dt\ d^3 x \left(\pi^{ab}\dot h_{ab}-N\mathcal{H}-N^a \mathcal{H}_a\right),\nonumber\\
\label{Einstein-Hilbert}
\end{eqnarray}
where $\mathcal{H}$ denotes the part of the Hamiltonian density referring to the space-time direction being
orthogonal to the hypersurface, $\mathcal{H}_a$ denotes the part of the Hamiltonian density referring to
the hypersurface, $K_{ab}$ denotes the extrinsic curvature, which can be expressed as

\begin{equation}
K_{ab}=\frac{1}{2N}\left(\dot h_{ab}-D_a N_b-D_b N_a\right),
\end{equation}
$\pi^{ab}$ denotes the canonical conjugated quantity to $h_{ab}$ and is of the following shape:

\begin{equation}
\pi^{ab}=\frac{\partial \mathcal{L}}{\partial \dot h_{ab}}=\frac{\sqrt{h}}{16 \pi G}\left(K^{ab}-K h^{ab}\right),
\label{canonical_conjugated_momentum}
\end{equation}
and $G_{abcd}$ is defined according to

\begin{equation}
G_{abcd}=\frac{1}{2\sqrt{h}}\left(h_{ac}h_{bd}+h_{ad}h_{bc}-h_{ab}h_{cd}\right).
\end{equation}
The quantization of the gravitational field,

\begin{equation}
h_{ab}\rightarrow \hat h_{ab}\quad,\quad \pi^{ab}\rightarrow \hat \pi^{ab},
\end{equation}
is performed in analogy to quantum mechanics by postulating commutation relations, in this case between the
three metric $h_{ab}$ and its canonical conjugated quantity $\pi^{ab}$. If there is presupposed a generalized
quantization principle as fundamental property of quantum theory, then with respect to the quantization of
the gravitational field this means that instead of postulating the usual commutation relation \cite{Kiefer:2004},\cite{Kiefer:2008sw},\cite{Giulini:2006xi},

\begin{equation}
[\hat h_{ab}(x),\hat \pi^{cd}(y)]=\frac{i}{2}\delta(x-y)\left[\delta^c_a \delta^d_b+\delta^d_a \delta^c_b \right],
\label{usual_commutation_relation_gravity}
\end{equation}
one has to postulate an extended commutation relation for the gravitational field:

\begin{equation}
[\hat h_{ab}(x),\hat \pi^{cd}(y)]=i\delta(x-y)\left\{\frac{1}{2}\left[\delta^c_a \delta^d_b+\delta^d_a \delta^c_b\right]
\left[1+\lambda A(\hat h,\hat \pi)\right]+\kappa B^{cd}_{ab}(\hat h, \hat \pi)\right\},
\label{generalized_commutation_relation_gravity}
\end{equation}
where $A(\hat h,\hat \pi)$ and $B^{cd}_{ab}(\hat h, \hat \pi)$ denote two arbitrary functions depending on the three metric
and its canonical conjugated quantity and $\lambda$ and $\kappa$ still denote constants. This quantization principle 
($\ref{generalized_commutation_relation_gravity}$) is obtained from the analogous generalized quantization principle in
quantum mechanics ($\ref{generalized_algebra_quantum_mechanics}$). The analogy between the generalization of the quantization
principle of general relativity ($\ref{generalized_commutation_relation_gravity}$) and the generalization of the quantization
principle in quantum mechanics ($\ref{generalized_algebra_quantum_mechanics}$) is represented in the tabular below. In
principle there are conceivable various concrete manifestations of the generalized quantization principle of the gravitational
field ($\ref{generalized_commutation_relation_gravity}$) leading to different kinematics of the corresponding quantum theory of
gravity. In the following two subsections are explored two special manifestations. This means that the usual quantum
description of the gravitational field in quantum geometrodynamics is generalized by postulating
($\ref{generalized_commutation_relation_gravity}$) and presupposing two different pairs of special generalization functions
$A(\hat h,\hat \pi)$ and $B^{cd}_{ab}(\hat h,\hat \pi)$.\\
\\
\begin{tabular}{|c|c|c|}
\hline
& quantum mechanics & quantum gravity\\
\hline
usual &&\\
quantization& $[\hat x_{\mu},\hat p^{\nu}]=i\delta_{\mu}^{\nu}$&$[\hat h_{ab}(x),\hat \pi^{cd}(y)]=
i\delta(x,y)\delta_{(a}^c \delta_{b)}^d$\\principle&&\\
\hline
generalized &&\\
quantization&$[\hat x_\mu,\hat p^\nu]=i\left\{\delta_{\mu}^\nu\left[1+\lambda A(\hat x,\hat p)\right]
+\kappa B_{\mu}^{\nu}(\hat x,\hat p)\right\}$&$[\hat h_{ab}(x),\hat \pi^{cd}(y)]=i\delta(x,y)
\left\{\delta_{(a}^c \delta_{b)}^d\left[1+\lambda A(\hat h,\hat \pi)\right]
+\kappa B_{ab}^{cd}(\hat h, \hat \pi)\right\}$\\
principle&&\\
\hline
\end{tabular}

\section{Representation of Operators and Corresponding Wheeler-DeWitt Equations in two Special Cases}

\subsection{Generalized Quantization Principle Depending on Canonical Conjugated Quantity}

As first special manifestation of ($\ref{generalized_commutation_relation_gravity}$) shall be considered a generalized
quantization principle, where the commutator between the three metric and its canonical conjugated quantity depends on
the canonical conjugated quantity.
This case is the transfer of a generalized uncertainty principle in quantum mechanics leading to a minimal length.
Especially it is presupposed the following pair of functions: $A(\hat h,\hat \pi)=\hat \pi^{mn}\hat \pi_{mn}$ and
$B^{cd}_{ab}(\hat h,\hat \pi)=2\hat \pi^{ab}\hat \pi_{cd}$, leading to the following shape of the postulated commutator:

\begin{equation}
\left[\hat h_{ab}(x),\hat \pi^{cd}(y)\right]=i\delta(x-y)\left[\frac{1}{2}\left(\delta^c_a \delta^d_b+\delta^d_a \delta^c_b \right)\left(1+\lambda \hat \pi^{mn}\hat \pi_{mn}\right)+2\lambda \hat \pi^{ab}\hat \pi_{cd}\right],
\label{generalized_quantization_conjugated}
\end{equation}
where has been set: $\kappa=\lambda$. The parameter $\lambda$ determining the influence of the additional terms of the generalized quantization principle is assumed to be proportional to the squared Planck length in case
of quantum mechanics. This leads to a corresponding minimal length, which is equal to the
Planck length. To maintain the analogy of the generalized quantization principle in case
of general relativity, one should assume that the parameter $\lambda$ is also directly related
to the squared Planck length. This means that it is assumed to be very small and accordingly
leads to the possibility to consider a series expansion to the first order.
This generalized commutation relation ($\ref{generalized_quantization_conjugated}$) is the analogue 
to ($\ref{special_generalied_algebra_quantum_mechanics}$) and can be realized by the following
representation of the operators $\hat h_{ab}$ and $\hat \pi^{ab}$ in the three metric space:
 
\begin{eqnarray}
\hat h_{ab}|\Psi[h]\rangle&=&h_{ab}|\Psi[h]\rangle,\nonumber\\
\hat \pi^{ab}|\Psi[h]\rangle&=&-i\left(1-\lambda \frac{\delta}{\delta h_{mn}}\frac{\delta}{\delta h^{mn}}\right)
\frac{\delta}{\delta h_{ab}}|\Psi[h]\rangle+\mathcal{O}\left(\lambda^2\right),
\label{metric_representation_conjugated}
\end{eqnarray}
which is analogous to ($\ref{generalized_operators_quantum_mechanics_position}$) and where has accordingly to be
used a series expansion of the operator $\hat \pi^{ab}$ in $\lambda$. In the representation referring to the canonical
conjugated quantity $\hat \pi^{ab}$ commutation relation ($\ref{generalized_quantization_conjugated}$) can be
realized as follows:

\begin{eqnarray}
\hat h_{ab}|\Psi[\pi]\rangle&=&i\left[\left(1+\lambda \pi^{mn}\pi_{mn}\right)\frac{\delta}{\delta \pi^{ab}}
+2\lambda \pi_{ab}\pi^{mn}\frac{\delta}{\delta \pi^{mn}}\right]|\Psi[\pi]\rangle,\nonumber\\
\hat \pi^{ab}|\Psi[\pi]\rangle&=&\pi^{ab}|\Psi[\pi]\rangle,
\label{conjugated_representation_conjugated}
\end{eqnarray}
which is analogous to ($\ref{generalized_operators_quantum_mechanics_momentum}$).
From ($\ref{conjugated_representation_conjugated}$) one can derive that the several components of the three metric
fulfil nontrivial commutation relations as well, if there is presupposed ($\ref{generalized_quantization_conjugated}$),
which are given by

\begin{equation}
[\hat h_{ab},\hat h_{cd}]=4\lambda^2 \pi^{mn} \pi_{mn}\left(\pi_{cd}\frac{\delta}{\delta \pi^{ab}}
-\pi_{ab}\frac{\delta}{\delta \pi^{cd}}\right).
\label{commutation_relation_metric}
\end{equation}
In \cite{Kober:2011am} has been presupposed a nontrivial noncommutativity algebra between the components
of the tetrad field and there have been derived generalized dynamics for general relativity by using coherent states.
($\ref{commutation_relation_metric}$) shows that such a structure could indeed arise from a quantum description of the gravitational field, if there is presupposed a generalized quantization principle.
The uncertainty relation between the three metric and its canonical conjugated quantity corresponding to
the generalized quantization principle ($\ref{generalized_quantization_conjugated}$) and being analogue
to the case in quantum mechanics ($\ref{generalized_uncertainty_quantum_mechanics}$) is given by

\begin{eqnarray}
\Delta h_{ab} \Delta \pi^{ab}&=&\frac{1}{2}|\langle \Psi|[h,\pi]|\Psi\rangle|
=\frac{1}{4}\left(1+3\lambda\langle\hat \pi^{ab}\rangle\langle\hat \pi^{ab}\rangle
+3\lambda\Delta \hat \pi_{ab}\Delta \hat \pi^{ab}\right)\quad {\rm if}
\quad a=c \wedge b=d\quad{\rm or}\quad a=d \wedge b=c,\nonumber\\
\Delta h_{ab} \Delta \pi^{cd}&=&\frac{1}{2}|\langle \Psi|[h,\pi]|\Psi\rangle|
=\lambda \langle\hat \pi_{ab}\rangle \langle\hat \pi^{cd}\rangle\quad {\rm if}
\quad a \neq c \vee b \neq d\quad {\rm and}\quad a \neq d \vee b \neq c.
\end{eqnarray}

\subsection{Generalized Quantization Principle Depending on Three Metric}

As second special manifestation of ($\ref{generalized_commutation_relation_gravity}$) shall be considered
a generalized quantization principle for the gravitational field depending on the three metric. Especially,
it is presupposed the following pair of functions: $A(\hat h,\hat \pi)=\hat h^{mn}\hat h_{mn}$ and
$B^{cd}_{ab}(\hat h,\hat \pi)=0$, leading to the following shape of the postulated commutator:

\begin{equation}
\left[\hat h_{ab}(x),\hat \pi^{cd}(y)\right]=i\delta(x-y)\left[\frac{1}{2}\left(\delta_a^c \delta_b^d
+\delta_a^d \delta_b^c \right)\left(1+\lambda \hat h^{mn}\hat h_{mn}\right)\right].
\label{generalized_quantization_metric}
\end{equation}
In the three metric space the quantization principle ($\ref{generalized_quantization_metric}$) can be realized
by the following representation of the operators $\hat h_{ab}$ and $\hat \pi^{ab}$:

\begin{eqnarray}
\hat h_{ab}|\Psi[h]\rangle&=&h_{ab}|\Psi[h]\rangle,\nonumber\\
\hat \pi^{ab}|\Psi[h]\rangle&=&-i\left(1+\lambda h^{mn}h_{mn}\right)
\frac{\delta}{\delta h_{ab}}|\Psi[h]\rangle.
\label{metric_representation_metric}
\end{eqnarray}
The representation referring to the canonical conjugated quantity will be omitted here. From
($\ref{metric_representation_metric}$) one can derive that the several components of the canonical
conjugated quantity fulfil the following commutation relations, if there is presupposed
($\ref{generalized_quantization_metric}$):

\begin{equation}
[\hat \pi^{ab},\hat \pi^{cd}]=2i\lambda\left(\hat h^{cd}\hat \pi^{ab}-\hat h^{ab}\hat \pi^{cd}\right).
\end{equation}
The generalized uncertainty relation between the three metric and its canonical conjugated quantity corresponding
to ($\ref{generalized_quantization_metric}$) is given by

\begin{eqnarray}
\Delta h_{ab} \Delta \pi^{cd}&=&\frac{1}{2}|\langle \Psi|[h^{ab},\pi_{cd}]|\Psi\rangle|
=\frac{1}{4}\left(\delta^c_a \delta^d_b+\delta^d_a \delta^c_b \right)
\left(1+\lambda \langle \hat h^{mn}\rangle \langle \hat h_{mn}\rangle+\lambda \Delta h^{mn}\Delta h_{mn}\right).
\end{eqnarray}
The corresponding shape of the quantum constraints and thus the Wheeler-DeWitt equations describing the dynamics
of the quantum states referring to the gravitational field with respect to the two special quantization conditions
is provided in the next subsection.

\subsection{Corresponding Generalized Wheeler-DeWitt Equations}

The quantum states for the gravitational field, $|\Psi[h]\rangle$, which are valid, are restricted by quantum
constraints being obtained by quantizing the Hamiltonian density of general relativity. The complete Hamiltonian
density of general relativity, $\mathcal{H}^{g}=N\mathcal{H}+N^a \mathcal{H}^{a}$, reads

\begin{equation}
\mathcal{H}^g=\pi^{ab}\dot h_{ab}-\mathcal{L}^g=16 \pi G N G_{abcd} \pi^{ab}\pi^{cd}
-\frac{\sqrt{h}}{16 \pi G}\left(R_h-2 \Lambda\right)-2 N_b D_a \pi^{ab},
\label{Hamiltonian_density}
\end{equation}
where $D_a$ denotes the covariant derivative with respect to the gravitational field. $\pi^{ab}$ and
$\mathcal{L}^g$ are defined in ($\ref{Einstein-Hilbert}$) and ($\ref{canonical_conjugated_momentum}$).
Variation of the Einstein Hilbert action ($\ref{Einstein-Hilbert}$) with respect to $N$ and $N_a$
yields the classical constraints

\begin{eqnarray}
\mathcal{H}=16 \pi G G_{abcd} \pi^{ab}\pi^{cd}-\frac{\sqrt{h}}{16 \pi G}\left(R_h-2 \Lambda\right)=0\quad,\quad
\mathcal{H}_a=-2 D_b  \pi_a^b=0.
\label{constraints}
\end{eqnarray}
The first equation of ($\ref{constraints}$) describes the Hamiltonian constraint and the second equation describes the
diffeomorphism constraint. To obtain the quantum constraints of the first quantization condition
($\ref{generalized_quantization_conjugated}$) there has to be inserted the corresponding representation of the operators with
respect to the three metric ($\ref{metric_representation_conjugated}$) to the constraints ($\ref{constraints}$) leading to

\begin{eqnarray}
\left\{16 \pi G G_{abcd}\left[\frac{\delta}{\delta h_{ab}}\frac{\delta}{\delta h_{cd}}
-2\lambda \frac{\delta}{\delta h_{mn}}\frac{\delta}{\delta h^{mn}}\frac{\delta}{\delta h_{ab}}\frac{\delta}{\delta h_{cd}}\right]+\frac{\sqrt{h}}{16 \pi G}\left(R_h-2\Lambda\right)\right\}|\Psi[h]\rangle
+\mathcal{O}\left(\lambda^2\right)=0,\nonumber\\
2i D_b h_{ac}\left(1-\lambda\frac{\delta}{\delta h_{mn}}\frac{\delta}{\delta h^{mn}}\right)
\frac{\delta}{\delta h_{bc}}|\Psi[h]\rangle+\mathcal{O}\left(\lambda^2\right)=0,
\label{quantum_constraints_generalized_conjugated}
\end{eqnarray}
if there used an expansion to the first order in the parameter $\lambda$. In case of the second quantization condition there has to be performed no series expansion. Here accordingly there has to be inserted the three metric representation of the
second case ($\ref{metric_representation_metric}$) to the constraints ($\ref{constraints}$) leading to the quantum constraints

\begin{eqnarray}
\left\{16 \pi G G_{abcd}\left(1+\lambda h^{kl}h_{kl}\right)
\left[\left(1+\lambda h^{mn}h_{mn}\right)\frac{\delta}{\delta h_{ab}}\frac{\delta}{\delta h_{cd}}
+2\lambda h^{ab}\frac{\delta}{\delta h_{cd}}\right]
+\frac{\sqrt{h}}{16 \pi G}\left(R_h-2 \Lambda\right)\right\}|\Psi[h]\rangle=0,\nonumber\\
2i D_b h_{ac}\left(1+\lambda h^{mn}h_{mn}\right)\frac{\delta}{\delta h_{bc}}|\Psi[h]\rangle=0.
\label{quantum_constraints_generalized_metric}
\end{eqnarray}
The first equations of ($\ref{quantum_constraints_generalized_conjugated}$) and ($\ref{quantum_constraints_generalized_metric}$) are the corresponding generalized Wheeler-DeWitt equations to
the quantization principles ($\ref{generalized_quantization_conjugated}$) and ($\ref{generalized_quantization_metric}$)
and thus describe the generalized quantum theoretical dynamics of the gravitational field.
In ($\ref{quantum_constraints_generalized_conjugated}$) and ($\ref{quantum_constraints_generalized_metric}$) has been
used a special choice of factor ordering corresponding to the transition:

\begin{equation}
G_{abcd} \pi^{ab} \pi^{cd} \longrightarrow \left[\frac{1}{2\sqrt{\hat h}}\left(\hat h_{ac}\hat h_{bd}
+\hat h_{ad}\hat h_{bc}-\hat h_{ab}\hat h_{cd}\right)\hat \pi^{ab} \hat \pi^{cd}\right].
\end{equation}

\section{Application to Quantum Cosmology}

Cosmological considerations in the context of a generalized uncertainty principle in quantum mechanics have
already been treated. Usual Cosmology with a minimal length has been treated in \cite{Ashoorioon:2004vm},\cite{Ashoorioon:2004rs},\cite{Ashoorioon:2004wd},\cite{Ashoorioon:2005ep},\cite{Vakili:2008tt},\cite{Garattini:2010dn}. Quantum cosmology under incorporation of a generalized uncertainty principle has been
considered in \cite{Battisti:2008rv}. In this section the general theory of quantum gravity with a generalized
quantization principle for the gravitational field developed in the last two sections shall be applied to
quantum cosmology. This means that the generalized quantization principle for the gravitational field also
leads to a generalized quantization rule for the scale factor, which represents the degree of freedom of
the metric in the cosmological case.
A similar quantization of the cosmological degree of freedom of the gravitational field can be found in
\cite{Majumder:2011ad}. But there has been treated a completely different scenario and as already mentioned
in the introduction the quantization of the cosmological degree of freedom of the gravitational field according
to a generalized uncertainty principle in \cite{Majumder:2011ad} seems not be interpreted as a special manifestation
of a generalized quantization principle for the gravitational field considered as a fundamental property
of quantum gravity.

\subsection{Derivation of the Generalized Cosmological Wheeler-DeWitt Equation}

In this section the generalized quantization condition for the gravitational field and the corresponding
Wheeler-DeWitt equation will be applied to cosmology. Especially it will be presupposed the second special
case ($\ref{generalized_quantization_conjugated}$), where the commutation relation between the three metric and
its canonical conjugated quantity depends on the canonical conjugated quantity rather then on the three metric.
The usual line element, $ds^2=g_{\mu\nu} dx^\mu dx^\nu$, of a Friedmann Lemaitre universe takes the following
special form:

\begin{equation}
ds^2=-N^2(t)dt^2+a^2(t)d\Omega_3^2,\quad{\rm with}\quad d\Omega_3^2=\frac{dr^2}{1-kr^2}
+r^2\left(d\theta^2+\sin^2 \theta d\varphi^2\right).
\end{equation} 
In this ansatz for the line element the three metric $h^{ab}$ is completely determined by the cosmological
scale factor $a$. If there is additionally incorporated a cosmological scalar field as matter field one can
obtain the following Hamiltonian as special manifestation of the Hamiltonian of quantum geometrodynamics
\cite{Kiefer:2004},\cite{Kiefer:2008sw}:

\begin{equation}
H=\frac{N}{2}\left(-\frac{\pi_a^2}{a}+\frac{\pi_\phi^2}{a^3}-a+\frac{\Lambda a^3}{3}+m^2 a^3 \phi^2\right),                  
\label{Hamiltonian_cosmology}
\end{equation}
if $\pi_a$ denotes the canonical conjugated quantity to the scale factor $a$ and $\pi_\phi$ denotes the canonical
conjugated quantity to the cosmological scalar field $\phi$. By integrating the two quantities $\pi_a$ and $\pi_\phi$
into a two dimensional vector space endowed with a metric $G_{AB}$ in the following way: 

\begin{equation}
\pi_A=\left(\pi_a,\pi_\phi\right)\quad,\quad G_{AB}=\left(\begin{matrix}-a&0\\0& a^3\end{matrix}\right),
\label{vector_space_metric}
\end{equation}
the cosmological Hamiltonian ($\ref{Hamiltonian_cosmology}$) can be written as

\begin{equation}
H=\frac{N}{2}\left(G^{AB}\pi_A \pi_B-a+\frac{\Lambda a^3}{3}+m^2 a^3 \phi^2\right).
\label{Hamiltonian_cosmology_rewritten}
\end{equation}
If the variables $a$ and $\phi$ and their canonical conjugated variables $\pi_a$ and $\pi_\phi$ shall
be quantized to obtain the corresponding Wheeler-DeWitt equation to the Hamiltonian
($\ref{Hamiltonian_cosmology_rewritten}$), there has to be performed the transition

\begin{equation}
G^{AB}\pi_A \pi_B \rightarrow \frac{1}{\sqrt{-G}}\hat \pi_A\left(\sqrt{-G}G^{AB}\hat \pi_B\right).
\label{transition_cosmology_general}
\end{equation}
The transition ($\ref{transition_cosmology_general}$) is based on a special choice of factor-ordering in analogy
to usuasl quantum cosmology.
In the usual case this would lead to $\frac{1}{a^2}\frac{\partial}{\partial a}\left(a \frac{\partial}{\partial a}\right)
-\frac{1}{a^3}\frac{\partial^2}{\partial \phi^2}$, since $\pi_a \rightarrow -i\frac{\partial}{\partial a}$
and $\pi_\phi \rightarrow -i\frac{\partial}{\partial \phi}$. In the scenario considered in this section
there is presupposed the generalized quantization principle ($\ref{generalized_quantization_conjugated}$)
concerning the gravitational field and thus there has to be performed another transition with respect to
the gravitational field, which is obtained from ($\ref{metric_representation_metric}$). The scalar field is 
however quantized in the usual way in this section, what can be considered as a kind of approximation.
This leads to the following transitions of the canonical conjugated quantities:

\begin{equation}
\pi_a \rightarrow \hat \pi_a=-i(1-\lambda\frac{\partial^2}{\partial a^2})
\frac{\partial}{\partial a}+\mathcal{O}(\lambda^2)\quad,\quad
\pi_\phi \rightarrow \hat \pi_\phi=-i\frac{\partial}{\partial \phi}.
\label{generalized_quantization_cosmology}
\end{equation}
Using ($\ref{generalized_quantization_cosmology}$) together with ($\ref{vector_space_metric}$)
in ($\ref{transition_cosmology_general}$) yields

\begin{equation}
-\frac{\pi_a^2}{a}+\frac{\pi_\phi^2}{a^3}=G^{AB}\pi_A \pi_B \rightarrow \frac{1}{a^2}\left(1-\lambda \frac{\partial^2}
{\partial a^2}\right)\frac{\partial}{\partial a}\left[a\left(1-\lambda \frac{\partial^2}{\partial a^2}\right)
\frac{\partial}{\partial a}\right]-\frac{1}{a^3}\frac{\partial^2}{\partial \phi^2}+\mathcal{O}\left(\lambda^2\right).
\label{transition_cosmology_special}
\end{equation}
This leads to the following expression for the cosmological Wheeler-DeWitt equation:

\begin{equation}
\frac{N}{2}\left\{\frac{1}{a^2}\left(1-\lambda \frac{\partial^2}{\partial a^2}\right)\frac{\partial}{\partial a}\left[a\left(1-\lambda \frac{\partial^2}{\partial a^2}\right)
\frac{\partial}{\partial a}\right]-\frac{1}{a^3}\frac{\partial^2}{\partial \phi^2}
+V(\phi,a)\right\}\Psi(a,\phi)+\mathcal{O}\left(\lambda^2\right)=0,
\label{WheelerDeWitt_cosmological}
\end{equation}
where it has been defined 

\begin{equation}
V(\phi,a)=-a+\frac{\Lambda a^3}{3}+m^2 a^3 \phi^2.
\end{equation}
The cosmological Wheeler-DeWitt equation ($\ref{WheelerDeWitt_cosmological}$) can be transformed to 

\begin{equation}
\left\{-\frac{2\lambda}{a}\frac{\partial^4}{\partial a^4}-\frac{4\lambda}{a^2}\frac{\partial^3}{\partial a^3}
+\frac{1}{a}\frac{\partial^2}{\partial a^2}+\frac{1}{a^2}\frac{\partial}{\partial a}
-\frac{1}{a^3}\frac{\partial^2}{\partial \phi^2}
+V(\phi,a)\right\}\Psi(a,\phi)+\mathcal{O}\left(\lambda^2\right)=0.
\label{WheelerDeWitt_cosmological_transformed}
\end{equation}

\subsection{Solution of the Generalized Cosmological Wheeler-DeWitt Equation}

Concerning the solution of the generalized cosmological Wheeler-DeWitt equation ($\ref{WheelerDeWitt_cosmological_transformed}$) it will be assumed that $m=0$ implying that the intersection
term between $a$ and $\phi$ in the potential $V(a,\phi)$ vanishes: $m^2 a^3 \phi^2=0$. Under this precondition
the two variables $a$ and $\phi$ can be separated meaning that the wave function can be expressed as a product
of wave functions only depending on one of the variables: $\Psi(a,\phi)=\psi(a)\varphi(\phi)$.
Inserting this separation ansatz to the cosmological Wheeler-DeWitt equation derived above ($\ref{WheelerDeWitt_cosmological_transformed}$) under the condition $m=0$ leads to

\begin{equation}
\frac{a^3}{\psi(a)}\left\{-\frac{2\lambda}{a}\frac{\partial^4}{\partial a^4}
-\frac{4\lambda}{a^2}\frac{\partial^3}{\partial a^3}
+\frac{1}{a}\frac{\partial^2}{\partial a^2}+\frac{1}{a^2}\frac{\partial}{\partial a}
+V(a)\right\}\psi(a)=\frac{1}{\varphi(\phi)}\frac{\partial^2 \varphi(\phi)}{\partial \phi^2}.
\end{equation}
Since the two sides of equation ($\ref{WheelerDeWitt_cosmological_separation}$) are equal, it has to
be assumed that both sides are equal to a constant leading to the following system of two equations:

\begin{eqnarray}
\left\{-\frac{2\lambda}{a}\frac{\partial^4}{\partial a^4}-\frac{4\lambda}{a^2}\frac{\partial^3}{\partial a^3}
+\frac{1}{a}\frac{\partial^2}{\partial a^2}+\frac{1}{a^2}\frac{\partial}{\partial a}
+V(a)+\frac{c}{a^3}\right\}\psi(a)&=&0,\nonumber\\
\frac{\partial^2 \varphi(\phi)}{\partial \phi^2}+c\varphi(\phi)&=&0.
\label{WheelerDeWitt_cosmological_separation}
\end{eqnarray}
Remember that in this whole section the first order approximation in the parameter $\lambda$ is presupposed. 
The solution of the equation referring to the scalar field $\phi$ is trivial: $\varphi(\phi)=\exp\left(i\sqrt{c}\phi\right)$.
To solve the equation referring to the scale factor $a$ there is used Sommerfelds polynomial method. Accordingly
it is assumed the following ansatz, where $\psi(a)$ is expressed as a general polynomial and thus reads

\begin{equation}
\psi(a)=\sum_{n=0}^{\infty}\chi_n a^n,
\label{polynom_ansatz}
\end{equation}
where the $\chi_n$ are arbitrary coefficients, which are determined in the following solution procedure.
The derivations of ($\ref{polynom_ansatz}$) are given by

\begin{eqnarray}
&&\frac{\partial \psi(a)}{\partial a}=\sum_{n=0}^{\infty}n \chi_n a^{n-1}\quad,\quad
\frac{\partial^2 \psi(a)}{\partial a^2}=\sum_{n=0}^{\infty}n(n-1)\chi_n a^{n-2},
\nonumber\\
&&\frac{\partial^3 \psi(a)}{\partial a^3}=\sum_{n=0}^{\infty}n(n-1)(n-2)\chi_n a^{n-3}\quad,\quad
\frac{\partial^4 \psi(a)}{\partial a^4}=\sum_{n=0}^{\infty}n(n-1)(n-2)(n-3)\chi_n a^{n-4}.
\label{derivatives_polynom_function}
\end{eqnarray}
Inserting ($\ref{polynom_ansatz}$) and ($\ref{derivatives_polynom_function}$) to the equation of ($\ref{WheelerDeWitt_cosmological_separation}$) referring to $a$ yields

\begin{eqnarray}
&&\sum_{n=0}^{\infty}\left[-2\lambda n(n-1)(n-2)(n-3)\chi_n a^{n-5}
-4\lambda n(n-1)(n-2)\chi_n a^{n-5}\right.\nonumber\\&&\left.
+n(n-1)\chi_n a^{n-3}+n \chi_n a^{n-3}-\chi_n a^{n+1}
+\frac{\Lambda}{3}\chi_n a^{n+3}+c\chi_n a^{n-3}\right]=0.
\end{eqnarray}
This equation can be transformed to

\begin{eqnarray}
&&\sum_{n=0}^{\infty}\left[-2\lambda(n+5)(n+4)(n+3)(n+2)\chi_{n+5}-4\lambda(n+5)(n+4)(n+3)\chi_{n+5}
\right.\nonumber\\&&\left.
+(n+3)(n+2)\chi_{n+3}+(n+3)\chi_{n+3}-\chi_{n-1}+\frac{\Lambda}{3}\chi_{n-3}+c\chi_{n+3}\right]a^n=0.
\end{eqnarray}
Reordering of the terms yields

\begin{eqnarray}
&&\sum_{n=0}^{\infty}\left\{\lambda\left[-2(n+5)(n+4)(n+3)(n+2)-4(n+5)(n+4)(n+3)\right]\chi_{n+5}
\right.\nonumber\\&&\left.
+\left[(n+3)(n+2)+(n+3)+c\right]\chi_{n+3}-\chi_{n-1}+\frac{\Lambda}{3}\chi_{n-3}\right\}a^n=0.
\end{eqnarray}
If this equation shall be fulfilled, then one is led to the following condition for the coefficients
of the polynomial:

\begin{eqnarray}
\chi_{n+5}=\frac{\left[(n+3)(n+2)+(n+3)+c\right]\chi_{n+3}-\chi_{n-1}+\frac{\Lambda}{3}\chi_{n-3}}
{\lambda\left[-2(n+5)(n+4)(n+3)(n+2)-4(n+5)(n+4)(n+3)\right]}.
\end{eqnarray}
By shifting the indices this recursion relation can be rewritten to 

\begin{eqnarray}
\chi_{n+2}=\frac{\left[n(n-1)+n+c\right]\chi_{n}-\chi_{n-4}+\frac{\Lambda}{3}\chi_{n-6}}
{\lambda\left[-2(n+2)(n+1)n(n-1)-4(n+2)(n+1)n\right]}.
\label{relation_coefficients}
\end{eqnarray}
It has now to be shown that the corresponding polynomial with the coefficients obeying ($\ref{relation_coefficients}$)
converges to maintain its normalizability, which is the precondition to be considered as physical solution.
Therefore the behaviour of the relation ($\ref{relation_coefficients}$) has to be explored for $n \to \infty$.
To study the case  $n \to \infty$ just the terms of highest order in $n$ remain important in the nominator
as well as the denominator leading to

\begin{eqnarray}
\lim_{n \to \infty}\chi_{n+2}&=&\lim_{n \to \infty}
\frac{\left[n(n-1)+n+c\right]\chi_{n}-\chi_{n-4}+\frac{\Lambda}{3}\chi_{n-6}}
{\lambda\left[-2(n+2)(n+1)n(n-1)-4(n+2)(n+1)n\right]}=\frac{n^2\chi_{n}}{-2\lambda n^4}=\frac{\chi_{n}}{-2\lambda n^2}.
\end{eqnarray}
This means that for $n \to \infty$ it holds

\begin{equation}
\chi_{n+2}=-\frac{\chi_{n}}{2\lambda n^2}.
\end{equation}
In case of the exponential function, $\exp(-x^2)=\sum_{n=0}^{\infty}\alpha_n x^n$, the coefficients behave as
$\alpha_{n+2}=\frac{-2\alpha_{n}}{n}$ for $n \to \infty$. Thus it is obvious that the polynomial corresponding
to ($\ref{relation_coefficients}$) converges, is normalizable and thus can be considered as physical solution.
The first coefficients the formula refers to are degrees of freedom, which have to be chosen at the beginning.
Since the formula ($\ref{relation_coefficients}$) relates coefficients, which exponent differs by two, the
coefficients referring to even exponents are independent of the coefficients to odd exponents. For $\chi_0$,
$\chi_1$ and $\chi_2$ the expression on the right hand side of ($\ref{relation_coefficients}$) is not defined,
since the denominator becomes zero as can be seen, if one chooses $n=-2$, $n=-1$ or $n=0$ leading to $\chi_0$,
$\chi_1$ or $\chi_2$ on the left hand side. $\chi_0$ and $\chi_1$ would have to be chosen even if the denominator
would not become zero, since it is a precondition of ($\ref{relation_coefficients}$) to be defined for $\chi_{n+2}$
that at least the coefficient $\chi_n$ is defined. The other coefficients, which take negative indices for the first
terms, have to be set to zero. There will now be determined the first ten terms of the solution polynomial, if there
is defined $\chi_0=\alpha$, $\chi_1=\beta$ and $\chi_3=\gamma$:

\begin{eqnarray}
&&\chi_0=\alpha,\quad
\chi_1=\beta,\quad
\chi_2=\gamma,\quad
\chi_3=\frac{(1+c)\beta}{-24\lambda},\quad
\chi_4=\frac{(4+c)\gamma}{-144\lambda},\quad
\chi_5=\frac{(9+c)(1+c)\beta}{11520 \lambda^2},\quad
\chi_6=\frac{(16+c)(4+c)\gamma-\alpha}{172800 \lambda^2},\quad\nonumber\\
&&\chi_7=\frac{(25+c)(9+c)(1+c)\beta-\beta}{-29030400\lambda^3},\quad
\chi_8=\frac{(36+c)[(16+c)(4+c)\gamma-\alpha]-\gamma+\frac{\Lambda}{3}\alpha}{-812851200 \lambda^3},\quad\nonumber\\
&&\chi_9=\frac{(49+c)[(25+c)(9+c)(1+c)\beta-\beta]-\frac{(1+c)\beta}{-24\lambda}+\frac{\Lambda}{3}\beta}{234101145600\lambda^4}.
\nonumber\\
\end{eqnarray}
This leads to the following solution of the generalized cosmological Wheeler-DeWitt equation ($\ref{WheelerDeWitt_cosmological_transformed}$):

\begin{eqnarray}      
\Psi(a,\phi)&=&\left\{\alpha+\beta a+\gamma a^2+\frac{(1+c)\beta}{-24\lambda} a^3
+\frac{(4+c)\gamma}{-144\lambda} a^4+\frac{(9+c)(1+c)\beta}{11520 \lambda^2} a^5
+\frac{(16+c)(4+c)\gamma-\alpha}{172800 \lambda^2} a^6\right.\nonumber\\&&\left.
+\frac{(25+c)(9+c)(1+c)\beta-\beta}{-29030400\lambda^3} a^7
+\frac{(36+c)[(16+c)(4+c)\gamma-\alpha]-\gamma+\frac{\Lambda}{3}\alpha}{-812851200 \lambda^3}a^8
\right.\nonumber\\&&\left.
+\frac{(49+c)[(25+c)(9+c)(1+c)\beta-\beta]-\frac{(1+c)\beta}{-24\lambda}+\frac{\Lambda}{3}\beta}{234101145600\lambda^4}a^9
+\mathcal{O}\left(a^{10}\right)\right\}\exp\left(i\sqrt{c}\phi\right).
\end{eqnarray}
The free parameters $\alpha$,$\beta$ and $\gamma$ as well as $c$ represent something like initial conditions
on the state $\Psi(a,\phi)$. Of course they have to be chosen in such a way that the state is normalized
and thus they are restricted. 

\section{Summary and Discussion}

In this paper a generalized quantization principle for the gravitational field in the canonical approach
to a quantum description of general relativity has been proposed in analogy to the generalized uncertainty
principle in quantum mechanics. There have been considered two special cases: In one case the generalized
quantization principle is assumed to depend on the canonical conjugated quantity to the three metric and in the
other case it is assumed to depend on the three metric itself. It have been considered the corresponding
representations of the operators describing the quantities of quantum geometrodynamics and the generalized
Wheeler-DeWitt equations describing the dynamics for both cases. After this there has been considered the
special application to quantum cosmology, where the quantization principle becomes manifest with respect to
the scale factor containing the degree of freedom of the metric in a Friedmann Lemaitre universe. By using
the representation of the operator of the scale factor there has been derived the cosmological Wheeler-De
Witt equation for the case of a generalized quantization principle depending on the canonical conjugated
quantity. This equation has been solved by using Sommerfelds polynomial method. Accordingly there has been
obtained a recursion formula for the coefficients by inserting a general polynomial depending on the
cosmological scale factor to the cosmological Wheeler-DeWitt equation. From this recursion relation can
be seen obviously that the corresponding polynomial converges and thus can be considered as physical solution.

If a generalized uncertainty principle in quantum mechanics or the closely related concept of noncommutative
geometry is considered as fundamental description of nature, then it seems to be natural to transfer the
corresponding generalized commutation relation to the quantization of quantities in other theories, especially
field theories, as it is usually done concerning the usual Heisenbergian algebra. Since the quantization
principle has to be applied to any classical theory to obtain the corresponding quantum theory, its
generalization also has to be assumed to be valid for all special theories including general relativity.
In this paper there has only been explored the basic idea of this concept and the corresponding
Wheeler-DeWitt equation has been solved for one special scenario. It could be promising to extend the considerations
given in this paper concerning the complete mathematical formalism of canonical quantum gravity and other special
applications. In principle the suggested generalized quantization principle for the variables within a quantum
description of the gravitational field can be presupposed in any special formulation of canonical quantum gravity.  
Accordingly the exploration of the generalized quantization principle for the gravitational field with respect to
Ashtekars variables and within loop quantum gravity based on these variables would be of great interest.

\end{document}